 \documentclass[letterpaper,  11pt, oneside]{article}
 \usepackage{graphicx}
  \title{Improved method of the Extensive Air Shower \\* arrival direction estimation}%
 \author{M.S.~Svanidze, Yu.G.~Verbetsky \\
 \small HEP,~~El.~Andronikashvili Institute of Physics,~Tbilisi,~0177,~Georgia.\\
 \small E-mail: \quad mananasvanidze@yahoo.com; \quad yuverbetsky@mail.ru.
        }%
 
\begin{document}
\maketitle
 \begin{abstract}
  The properties of measuring systems of the Extensive Air Showers
  arrival directions (EAS goniometers), consisting of some sets of
  widely separated detectors registering the moments of EAS frontal
  passage, are considered. The advantage of volumetric goniometers
  in comparison with the commonly used flat ones is shown. The
  proper points selection method for detectors spatial arranging is
  suggested, providing the best accessible  accuracy of the EAS
  arrival direction estimation within the given detectors' number
  and installation overall dimensions.
  \end{abstract}

 \section{Introduction}

 Since the discovery of cosmic rays (CR) the problem of large-scale
 correlations has been almost disregarded. This is because charged
 particles, composing a main part of CR, must be deflected in the
 galactic magnetic field throughout their long journeys to the
 Earth, resulting in random injections into atmosphere of CR particle pairs,
 coinciding both in time and in direction. However, there exist
 some processes, engendering such pairs and groups of correlated
 primary particles in cosmic radiation. There may be, e.g., groups
 from the $\gamma$-bursts, the product $\gamma$-rays from super
 high energy collisions in the interstellar substance (especially
 in the immediate vicinity of CR generation areas), or the products
 of CR nuclei disintegration in the solar photon field -- i.e. the
 effect presented by Gerasimova and Zatsepin \cite{GZ,MT}, etc.

 All this effects are very rare, but during the last years there
 appeared some observational evidence of their existence
 \cite{CM,Feg,Alex} . All of them were performed by
 observation of correlated Extensive Air Showers (EAS).
 Actually, only both the registration
times' coincidence and the coincidence of two EAS cores'
directions with account of spatial positions of registration
points on the Earth's surface, can state as a partial warranty of
their genetic relationship

The problem can be solved in principle by the pool of modern
complicated installations, investigating the $\gamma$-initiated
EAS, such as the MAGIC in the Canary Islands \cite{mag} or VERITAS
in southern Arizona \cite{ver}. These installations detect and log
out the showers by observation of Cherenkov light generated in the
atmosphere by the quick charged particles composing the EAS. The
arrival direction measurement precision of $0.15^\circ$ is
expected \cite{fvt} for individual $\gamma$-quanta.

However, these extremely perfect installations are rarely disposed
on the Globe. The active observation time is limited by the
atmosphere state condition. The viewing angle usually is moderate,
e.g. the VERITAS array \cite{fvt} observes concurrently the sky
area of $3.5^\circ$ only. Thus it is not expectable for these
expensive installations to constitute some effective pool for
uninterrupted monitoring of rare events of widely spaced
genetically related showers.

 It is safe to say that the widespread network of small
traditional ground-based installations, recording both the EAS
event absolute astronomical time (UT) and the shower arrival
direction on the local upper hemisphere (and measuring the EAS
energy too, if possible), would make up more suitable tool for the
problem. The installations of this type, being comparatively
simple and inexpensive, would form reasonable equipment for
university teems and research groups round the Globe, constituting
the synchronized united network.

Now, as a matter of notational convenience, let us term any
ground-based installation, intended for the EAS arrival direction
measurement, as \emph{EAS goniometer}.

      The LAAS group \cite{Ochi} has performed $\,$ the network observation $\,$ of EAS $\,$ with
      energy
 \mbox{$(10^{13} \div 10^{16})eV$,} using ten independent scintillator
 goniometers arranged over a very large area. Especially this
 last group has investigated the EAS pairs coincidence not only by
 the times registered (using the Global Positioning System GPS) at
 the network stations, but also by the EAS arrival directions
 coincidence.
      This group has declared the attained measurement accuracy of EAS arrival direction to be about
 $10^\circ$ for the zenith angles less then $45^\circ$. This
 attainment is too small for the reliable paternal affiliation in
 the pairs observed. The analysis of   the typical
 station construction of the network \cite{Ochi} gives the opportunity to
 ascertain the possibility of much greater angular accuracy
 achievement even with the same number of detectors as in the EAS
 goniometer used there. It becomes possible to get an acceptable
 accuracy even for big zenith angles.
 This accuracy growth is very important for
 any network of remote stations as far as the potentially
 correlated showers prove to be close to the horizon for big enough angular
 distances between the stations. Just for the last cases the
 extensively used flat EAS goniometers with poor accuracy at large
 zenith angles are especially objectionable.

 \section{Volumetric EAS goniometer. Common case.}
  Generally the set of any $N_0$ ionizing radiation detectors,
  arbitrarily distributed in the 3--space, can be
 used as the volumetric EAS goniometer. Some ``triggering
 structure'' for EAS discovering is implied. This structure has to
 send a trigger signal to the measuring part of installation to
 start a time reckoning of the pulses from EAS goniometer
 detectors. In particular, the EAS goniometer detectors themselves
 can be used for this purpose. The signals from all detectors have
 to be delayed for the common period, $t_{del}$, with respect to the real
 moments of the pulses origins. This delay period has to be such that at any possible case
 of the shower arrival direction all
 $N_0$ detector signals would hit the measuring part of
 installation later then the trigger signal. The measuring part
 itself records the differences of signal arrival times with respect to the
 trigger signal hit time as well as the absolute astronomic time
 (UT) of trigger signal obtained by means of GPS. Only this set of
 recorded numbers has to be used later (\textsf{off line}) to
 estimate the direction and the shower front passing time through
 the installation's coordinate system origin.

      Hereinafter all delay periods are measured in distance unites , that is to say their clock estimations are
 multiplied in advance by the EAS front velocity, which is taken
  to be approximately equal to the light velocity.

      Let us fix the rectangular coordinate system for the volumetric EAS goniometer. All coordinates, times,
 their differences and distances are referred hereinafter to the
 mentioned coordinate system and are measured in meters. Furthermore, let us
 designate the coordinates of used detector centers as
 $$
     \textbf{r}_i(x,y,z) \quad i=1,2,3\ldots N_0
 $$

      The EAS front plane equation \cite{Korn} at arbitrary moment is
 \begin{equation}\label{1}
     \textbf{r} \cdot \textbf{n}-p=0
 \end{equation}

      Here \textbf{r} is an arbitrary point in the front plane; vector
 \textbf{n} is the main unit  ort of the plane,
 $\textbf{n}^2=1$, while the $p$ parameter measures the distance
 between the plane and coordinate system origin. Our aim is to
 estimate the ort components by the measured times of the front
 passage through the detectors.

      The distances from the detectors to the front plane in any position, specified by $p$ parameter value, is
 determined by the linear relationship
 \begin{equation}\label{2}
   \delta_i=\textbf{r} \cdot \textbf{n}-p \quad i=1,2,3\ldots N_0
 \end{equation}

      Let us select from the whole family of planes, corresponding to the different moments of the shower
 propagation,  the unique plane containing the coordinate
 origin. The last restriction implies the selection of the special
 plane from the planes'
 family (\ref{1}), singled out by the condition $p = 0$ . Just this plane will be referred as ``the EAS
 front plane''.

      For this unique EAS front plane the set (\ref{2}) of detectors' distances from the front plane
      are
 \begin{equation}\label{3}
   \delta_i=\textbf{r} \cdot \textbf{n}
 \end{equation}

      There can be some negative distances between them in common case, as any of the detector points
      $\textbf{r}_i$
 can be disposed on any side the EAS front plane. That is why we
 shall add an artificial and big enough common delay period $t_{del}$
 to both sides of equation (\ref{3}) (for instance, this delay can
 be attained by use of coupling cables with identical and big
 enough lengths):
 \begin{equation}\label{4}
   \textbf{r} \cdot \textbf{n}+t_{del}=\delta_i+t_{del}\doteq t_{tr}+t_i
 \end{equation}

      Here $t_i$ values are the measured by the EAS goniometer installation delay periods of detectors' signals
 with respect to trigger signal (all of them are the positive
 values), while $t_{tr}$ is the (unknown) triggering time.

      The decision variables in equation (\ref{4}) are:
      the common difference of delay period  of the moment when the EAS
 front plane passes the coordinate system origin of the
 installation with the triggering time, $t_\delta=t_{del}-t_{tr}$, and three nondimensional components of main ort
 \textbf{n}, i.e. the directional cosines of EAS core. The common
 difference period $t_\delta$ permits to obtain the UT of EAS front plane
 pass through the coordinate system origin (with use of the
 GPS--measured triggering time and common signal delay period known).
 Since there are 4 decision variables, so the requirement on the
 detectors number $N_0 \geq 4$   is the solvability condition of
 \mbox{equation (\ref{4}).} It is possible to get the solution even for
 three detectors, but it proves to be the singular case of \mbox{system
 (\ref{4}),} demanding a special solution method. This case of
 ``flat'' EAS goniometer is considered later. Hereon the common
 case is studied.

      For the case of detector number $N_0 > 4$  the EAS goniometer equations system
 \begin{equation}\label{5}
   \textbf{r} \cdot \textbf{n}+t_\delta=t_i
 \end{equation}
 is overfilled. The Least-Squares Method (LSM) \cite{Cramer,RPP} is
 to be used to determine the solution.

      It is convenient to define new  vector ${\nu}$ with 4
      rows     for the generalized main ort of the shower front plane,
 with the nondimensional $4^{th}$ component ${\nu}_t$ to measure
 the value of desired common delay difference. The coordinates of all used
 detectors constitute the $4 \times N_0$ matrix ${\rho}$ with
 dummy $4^{th}$ column, containing identical $4^{th}$ component
 value $\lambda = 1m$ for all detectors. The measured delay periods
 constitute the $N_0$-row vector ${\tau}$. Now the equation
 (\ref{5}) takes the form of common matrix one:
 \begin{equation}\label{6}
     \rho\nu=\tau
 \end{equation}

      As usual \cite{Cramer,RPP} for the overfilled equation set in the frames of LSM, for delay periods distributed with
 dispersion matrix ${\Sigma}$, let us apply the left multiplication
 by $({\rho}^T \Sigma^{-1})$ matrix to the \mbox{equation
 (\ref{6})}. This results in normal equations set for the 4
 decision components of the generalized direction ort of the EAS
 front plane
 \begin{equation}\label{7}
     A\nu=d
 \end{equation}
      with the symmetric $4 \times 4$ square matrix
 \begin{equation}\label{8}
     A={\rho}^T \Sigma^{-1} {\rho}
 \end{equation}
      and new right-hand member of equation
 \begin{equation}\label{9}
     d={\rho}^T \Sigma^{-1} {\tau}
 \end{equation}

      It has the unique solution for the nonsingular matrix ${A}$. Just due to (\ref{8}) it becomes necessary to arrange
 the detectors set in full 3D space, not in any plane in it. The
 explicit solution is
 \begin{equation}\label{10}
     \nu=A^{-1} d
 \end{equation}
      or, directly expressed by the delay vector $\tau$
 \begin{equation}\label{11}
     \nu=G\tau
 \end{equation}
      Here ${G}$ is the $N_0 \times 4$ matrix
 \begin{equation}\label{12}
     G=A^{-1}\rho^T \Sigma^{-1}=(\rho^T \Sigma^{-1} \rho)^{-1} \rho^T \Sigma^{-1}
 \end{equation}

The estimation (\ref{11}) of the generalized main ort evaluated
here is consistent, unbiased and asymptotically normal estimator
(i.e. the expectation value of the estimator coincides with the
true value of the ort), as it is received from the initial data by
means of the linear Least Squares Method\cite{SMExPh}.

      Essentially, expression (\ref{11}) is the whole solution in the common      case,
      but the error analysis for this solution gives one the possibility to recognize some additional location
 requirements for the final achievement of desired estimation
 quality of EAS direction.

      At this stage let us assume the coordinates of
detectors to be taken as exactly prescribed.
 Practically it means that the detectors' location errors have to
 be less then the errors of delay periods  at least by an order of
 magnitude.

      The generalized direction ort, ${\nu}$, is linearly connected with the measured delay periods \mbox{by (\ref{11})}.
       So the dispersion matrix,  ${D}$ (of the direction ort ${\nu}$), is connected with
 the delay dispersion \mbox{matrix, ${\Sigma}$}, by the linear
 transformation \cite{RPP}, too:
 \begin{equation}\label{13}
     D=G \Sigma G^T
 \end{equation}

      At this stage let us assume as a hypothesis that all delay periods, $t_i$, are identically distributed
 independent quantities with the same dispersion values,
 ${\sigma}^2$. This assumption is very close to reality, indeed, as
 the processes in one delay device do not affect the properties of another
 one. Hence, there are no correlations in the dispersion matrix ${\Sigma}$ of the delay
vector $\tau$.  On the other
 hand, all delay errors in every signal path originate from the similar
 reasons, so they can be considered as equal on average. Therefore
 we can accept the relation
 \begin{equation}\label{14}
     \Sigma=\sigma^2 I
 \end{equation}

 Here ${{I}}$ is the $4 \times 4$ unity matrix.

      In this special case the dispersion matrix of the required vector ${\nu}$ can
 be expressed in explicit form via the ${{A}}$ matrix (\ref{7}):
 \begin{equation}\label{15}
     D=A^{-1}
 \end{equation}

      So, all volumetric EAS goniometer properties are determined through the detectors arranging
 matrix ${\rho}$ and delay dispersion value ${\sigma}^2$ (see
 (\ref{8})(\ref{14})).

      Broadly speaking, it is commonly desirable for the sought quantities (i.e. the generalized direction
 ort ${\nu}$ components in our case) to be statistically
 uncorrelated estimators at least. It is desired for the component
 dispersions to be equal, too. These requirements are strong
 enough. They will give us the possibility to define more
 accurately the best scheme of detectors arranging in the space.

 \section{Multitier EAS goniometer}

 \mbox{\qquad If} it is planed to place the EAS goniometer on the flat horizontal plain, the local vertical line becomes
 the preferential direction in the space and it is natural to
 orientate the installation upon the last one. It is clear that the
 error isotropy of EAS direction estimations for the upper
 hemisphere is desired.

\begin{figure}[t]
\begin{minipage}[c]{.4\linewidth}
 \begin{center}
 \parbox[t]{6cm}
  {\small
      { \textsf{Detectors are represented by points in the
       polygon apexes at any level. The detector numbers for the example shown are:} $$N_1=8;\;N_2=8;\;\ldots N_M=4. $$
       \textsf{The planes of tiers are displayed conditionally for visual manifestation of axial symmetry only.}
      }
   }
 \end{center}
\end{minipage}%
\hspace{.05\linewidth}
\begin{minipage}[c]{.45\linewidth}
\centering
\includegraphics[natwidth=156mm,natheight=205mm,width=75mm]{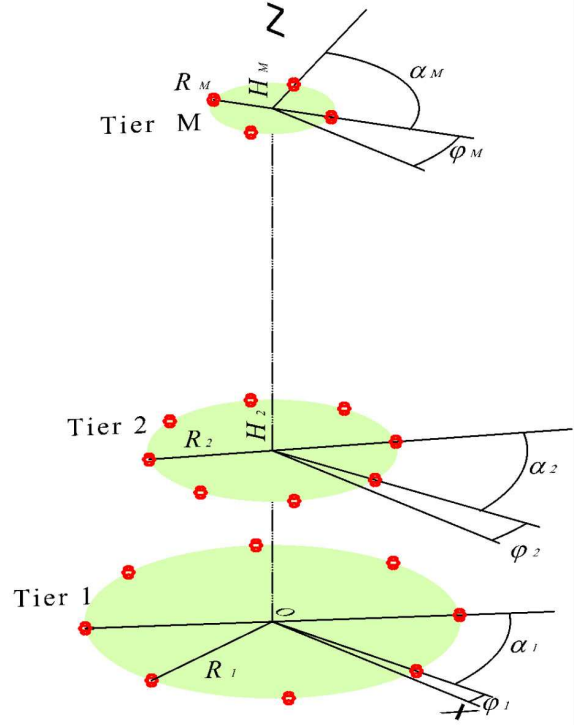}
\end{minipage}
\centering \caption{The detectors' position scheme for the
multitier EAS goniometer.} \label{schema_of_gonio} \vspace{0.5cm}
\end{figure}

  \mbox{\qquad Let} us consider a multitier scheme of volumetric EAS goniometer. It consists of $N_0$ detectors arranged
 on several horizontal plane levels (tiers). Such a construction
 prevents the possible singularity of equation (\ref{7}). It is
 clear that the desirable azimuth symmetry of errors results in
 axial symmetry of detectors position on every tier. In the
 simplest case ( see Fig.\ref{schema_of_gonio}) they are placed
 uniformly along the circumferences with centers based on a common
 vertical axis.

  \mbox{\qquad All} detectors hereon belong to the goniometer subsystem itself, the trigger subsystem is not
 considered.

      Let us use the rectangular frame of reference with horizontal $XOY$ plane and with $OZ$ axes directed
 upwards along the local vertical line. Axes $OX$ is a polar one
 for the azimuth angles.

      All detectors are situated on $M$ horizontal levels;
          $\quad a = 1, 2, 3 \ldots M \quad$   are indexes of these levels.

      Every $a$--level contains $N_a$ detectors. The total number of detectors is

 \begin{equation}\label{16}
     N_0=\sum_{a=1}^{M} N_a
 \end{equation}

      The radii of the detectors' positions regarding to the $OZ$ axis are $R_a$ at any a--level, while $H_a$ are the
 levels' heights above the $XOY$ plane.

      At any $a$--level a separate detectors numeration is established: $k = 1, 2, 3, \ldots N_a \quad$  are the indexes
      of detectors    at each level.

      These detectors divide uniformly corresponding circle of $R_a$ radius with the angular step
 \begin{equation}\label{17}
     \alpha_a=\frac{2\pi}{N_a}
 \end{equation}

      The phase shifts of detectors situated along the $a$--level circumferences are denoted as $\varphi_a$.
       Every delay period $t_{a,k}$ of detector with item number $k$, disposed on the $a$--level, has to be
 recorded as a result of installation triggering due to EAS front
 passage through the EAS goniometer.

      It is clear that the $a$--level detectors are situated in the points with the coordinates:
 \begin{equation}\label{18}
     (X_{a,k}=R_a \cos(k \alpha_a +\varphi_a),\; Y_{a,k}=R_a \sin(k \alpha_a
     +\varphi_a),\; H_a,\; \lambda)
 \end{equation}
      \begin{flushright}
      {--- just these sets of coordinates  constitute the $N_0$-–-row matrix $\rho$ of detectors'
      positions.}
      \end{flushright}

      The circles division mode used above provides the definitive calculation \cite{PBM} of \mbox{the $A$ matrix (\ref{8})} in the
 explicit form.

 It is convenient to define the ``tier averaging''
 operation for any set of values $Q_a$ representing some property
 of every level. We shall designate this operation by broken
 brackets:
 \begin{equation}\label{19}
     < Q >\quad \stackrel{\mathrm{def}}{=} \quad\frac{1}{N_0} \sum_{a=1}^{M} N_a Q_a
 \end{equation}

       The explicit form of matrix $A$  turns out to be
 \begin{equation}\label{20}
     A=\frac{N_0}{\sigma^2}\cdot  \left(%
 \begin{array}{cccc}
   \displaystyle \frac{< R^2 >}{2} & 0 & 0 & 0 \\
   0 & \displaystyle \frac{< R^2 >}{2} & 0 & 0 \\
   0 & 0 & \displaystyle < H^2 > & \lambda < H >  \\
   0 & 0 & \lambda < H > & \displaystyle \lambda^2 \vphantom{\frac{< R^2 >}{2}}\\
 \end{array}%
 \right)
 \end{equation}

      The matrix is independent of the phase shifts on every level \cite{PBM}.
      It proves to be a block--diagonal one with only nondiagonal elements corresponding to the \mbox{$3^{rd}$ (zenith)} and
 \mbox{$4^{th}$ (temporal)} components of the solution $\nu$. The
 dispersion matrix $D=A^{-1}$ has the same structure.

      It turns out than, that the correlation between the zenith and temporal
      components of the generalized ort $\nu$ estimation
      can be cancelled out by a simple selection of levels heights.
      These heights have to satisfy a relation:

 \begin{equation}\label{21}
     < H >\quad =\quad\frac{1}{N_0} \sum_{a=1}^{M} N_a H_a\quad=\quad 0
 \end{equation}

      Hence, some of them have to be negative, i.e. the origin of installation coordinate system must be
 located above some of the lower tiers. It requires only a simple
 vertical shift of the initial coordinate system. As long as the
 last operation is a mathematical one and \emph{does not require any
 technical change} in the installation, it can be accomplished
 \emph{at any circumstances}. After this shift the  matrix $A$ reduces into a diagonal one with new value of
 $<H^2>$. It means that the columns in the detectors' positions
 matrix $\rho$ become orthogonal.

 Hereon we shall suppose that the condition (\ref{21}) is fulfilled
 --- \emph{it doesn't cost anything! }
      It is easy now to present an explicit solution of normal LSM equation (\ref{7}),
      i.e. the estimation of the front plane
 generalized ort:
\begin{equation}\label{22}
    \begin{array}{rcl}
      \nu_x & = & 2\;{<RC>}/{<R^2>} \vphantom{\frac{x^2}{x^2}}\\
      \nu_y & = & 2\;{<RS>}/{<R^2>} \vphantom{\frac{x^2}{x^2}}\\
      \nu_z & = & \phantom{.\;}{<HT>}/{<H^2>} \vphantom{\frac{x^2}{x^2}}\\
      \nu_t & = & \phantom{22\;}{<T>}/{\phantom{1}\lambda}\vphantom{\frac{x^2}{x^2}}\\
    \end{array}
\end{equation}

 Here are used some ``tiered averaged'' values, evaluated both from
 measured delay periods of detector signals and the angular
 coordinates of the detectors:
\begin{equation}\label{23}
    \begin{array}{rcl}
      C_a & = & \displaystyle \frac{1}{N_a} \sum_{k\,=\,1}^{N_a}{t_{a,k}\;\cos(k\alpha_a+\varphi_a)} \vphantom{\sum_{k=N_a}^{N^{a^2}}{t}}\\
      S_a & = & \displaystyle \frac{1}{N_a} \sum_{k\,=\,1}^{N_a}{t_{a,k}\;\sin(k\alpha_a+\varphi_a)} \vphantom{\sum_{k=N_a}^{N^{a^2}}{t}}\\
      T_a & = & \displaystyle \frac{1}{N_a} \sum_{k\,=\,1}^{N_a}{t_{a,k}} \vphantom{\sum_{k=N_a}^{N^{a^2}}{t}}\\
    \end{array}
\end{equation}

The common delay periods' difference $\quad
t_\delta=t_{del}-t_{tr} \quad$ now can be calculated explicitly
through the definition:
\begin{equation}\label{24}
    t_\delta=\lambda\cdot\nu_t\;=\;<T>
\end{equation}
      i.e. it is simply an arithmetical mean of all delay periods of all detector signals in the EAS goniometer.

      The estimation of ort $\nu$ dispersion matrix $D$ is shown above for identically distributed independent
 delay periods of detectors' signals (\ref{15}) with the same
 dispersion values $\sigma^2$. In the case of ``orthogonal'' multitier
 installation (i.e. with diagonal $A$ matrix) it reads:
 \begin{equation}\label{25}
     D_x=D_y=\frac{2 \sigma^2}{N_0 < R^2 >};\quad D_z=\frac{\sigma^2}{N_0 < H^2
     >};\quad D_t=\frac{\sigma^2}{N_0 \lambda^2};
 \end{equation}

It is strongly desirable to achieve an isotropy of estimation
accuracy for all 3D--ort $\textbf{n}$ components, too. This aim
 can be reached if the radii and the heights of all tiers  would be fitted to
 satisfy the relation  $< R^2 > = 2< H^2 >$, following the $D$
 matrix explicit view (\ref{25}) for ``orthogonal'' goniometers. Hence,
 the values of tiers' heights would to be of the same order of
 magnitude as the radii used, though it may prove to be difficult
 for realization.

      According with LSM deductions, the true estimation of dispersion $\sigma^2$ of delay periods
      is the (corrected) average
 of squared residual differences:
 \begin{equation}\label{26}
     s^2=\frac{1}{(N_0 -
     4)}\sum_{a=1}^{M}\sum_{k=1}^{N_a}(X_{a,k}\nu_x+Y_{a,k}\nu_y+Z_{a,k}\nu_z+\lambda \nu_t -t_{a,k})^2
 \end{equation}

      Here the estimation of the front plane generalized ort (\ref{22})(\ref{23}),
      the detectors \mbox{coordinates (\ref{18})} and
 measured delay periods must be substituted. Thus the dispersion of
 measured direction can be estimated \emph{for every EAS event}, but the
 goniometer must contain strictly more than \mbox{4 detectors} (see
 (\ref{26})).

      The dispersion matrix $D =A^{-1}$ displays (\ref{20})(\ref{25}) some special features of the orthogonal EAS
 goniometer scheme under consideration:


       \begin{quote}
       {a) all possible correlations can be eliminated by a simple coordinate
      shift;}

      {b) the dispersion values of ``horizontal'' components of the ort estimated
  are equal and proportionate to \mbox{$1/( N_0 \; <R^2>)$;\,} the ``vertical''
  one proportionate to \mbox{$1/( N_0 \; <H^2>)$;}}

      {c) the dispersion value of difference of common delay periods
  does not depend on the installation overall dimensions and
  proportionate to $1/(N_0)$.}\end{quote}

      While the $3D\,-$~ort $\textbf{n}$ is estimated by means of foregoing procedure (the additional
      ``temporal''component will be out of consideration hereon!),   the
 problem arises of  the corresponding   spherical angles estimating for the EAS
 arrival direction, i.e. of azimuth \mbox{angle $\phi$} and of zenith
 \mbox{angle $\theta$.}

      The direction ort $\textbf{n}$ components are defined through these angles with standard relations:
 \begin{equation}\label{27}
     \begin{array}{rcl}
       n_x & = & \sin\theta \; \cos\phi \\
       n_y & = & \sin\theta \; \sin\phi  \\
       n_z & = & \cos\theta \\
     \end{array}
 \end{equation}

      It is obvious that $\textbf{n}$ vector ought to be of unit length.
      This condition may be violated exceptionally by
 the errors in the ort components estimations. Thus it is
 reasonable and handy to prefer the angles calculation method
 exploiting only the components ratios, as it excludes the
 influence of accidental length variation. So, we accept for
 computations the special form of solution \mbox{of (\ref{27}):}
 \begin{equation}\label{28}
     \begin{array}{rcl}
        \cot\phi & = & {\textstyle n_x}/{\textstyle n_y} \vphantom{\sqrt{\frac{n_x}{n_y}}} \\
         \tan\theta & = & \sqrt{ \left(\frac{\textstyle n_x}{\textstyle n_z}\right)^2+\left(\frac{\textstyle n_y}{\textstyle n_z}\right)^2} \\
     \end{array}
 \end{equation}

      The last are  nonlinear expressions, so the usual corrections \cite{RPP},
      depending on the estimations of the dispersion matrix $D$ components (\ref{25}),
      must be used for angles estimations.

      The dispersion matrix of the angles obtained is a function of the ort \textbf{n} components' dispersions,
 derived above (\ref{25}). Let us calculate the matrix of first
 derivatives of the angles (\ref{28}) upon the ort components
 $(n_x, n_y, n_z)$:
 \begin{equation}\label{29}
     F_3(\phi, \theta)= \frac{\partial(\phi, \theta)}{\partial(n_x, n_y, n_z)}
 \end{equation}

      Following the method of error propagation \cite{RPP}, the dispersion matrix of spherical angles is
 \begin{equation}\label{30}
     \left(%
 \begin{array}{cc}
   \sigma_\phi^2 & C_{\phi\theta} \\
   C_{\phi\theta} & \sigma_\theta^2 \\
 \end{array}%
 \right)= F_3^T(\phi, \theta) D_3 F_3(\phi, \theta)
 \end{equation}
      Here $D_3$ is the spatial part of the dispersion matrix (\ref{25}) for ort $\nu$ components' estimations.

      For the EAS goniometer with the axial symmetry it results in relations
 \begin{equation}\label{31}
     \sigma_\phi^2=\frac{D_{h}}{\sin^2\theta};\quad
     \sigma_\theta^2= D_{h}\cos^2\theta +
     D_{z}\sin^2\theta; \quad C_{\phi\theta}=0.
 \end{equation}
      Here $D_h=D_x=D_y.$

      In this axially symmetric case the $\{\phi\theta\}\, -\, $covariation vanishes, both dispersions
depend on the true zenith angle only. The fast increase of
dispersion of azimuth angle estimation is a direct sequence of
spherical coordinate system singularity: at the $\theta \to 0$
limit the azimuth angle value is fundamentally indefinite. The
zenith angle value is limited at any case.

\section{Flat EAS goniometer}

  \mbox{\qquad Let} us investigate the possibilities of flat goniometers by the method used. The only difference with
 3D case consists in mutual equality of the $3^{rd}$ coordinate of every
 detector: $Z_{a,k}  \equiv const$. Hence the equation system
 (\ref{7}) becomes singular; it contains no information about the
 $3^{rd}$ component of the front plane ort; the equations have no complete
 solution.

      The way out of the difficulty consists in rejection of the $3^{rd}$ component mentioning from the LSM
 equation system. Let us evaluate now the ``horizontal'' and
 ``temporal'' components only, erasing both $3^{rd}$ column and row
 out of the equation system (\ref{7}).
 So we can estimate the ``horizontal'' and ``temporal'' ort
 components only, being consistent,
 unbiased and asymptotically normal estimator, just as in common case (\ref{11}). This is sufficient for the
 azimuth angle estimation. The value of the $3^{rd}$ component can
 be reconstructed by a formal way from the unity condition for the
 3D-ort length:
 \begin{equation}\label{32}
     n_z(n_x,\;n_y)=\sqrt{1-(n_x^2+n_y^2)}
 \end{equation}

      In this case the $3^{rd}$ component is not the independently estimated value, but a nonlinear function of the
 estimations of the ``horizontal'' components. Except the last
 moment the solution of problem is similar to that obtained above for the
 common case.

 But expressions (\ref{22}) and (\ref{32}) only nominally solve the
 problem of determination of the shower
 direction spherical angles . The
difficulty results from the connection (\ref{32}),
 since the ``horizontal'' components are estimated quantities with
 random errors.

      Let us employ again the method of error propagation \cite{RPP} and calculate the matrix of first derivatives of
 the angles (\ref{28}). This time only two ``horizontal''
 components are independent variables.

       It turns out that the error of the zenith angle
 is strongly rising  \mbox{(as
 $\sigma^2_\theta={D_h}/cos^6(\theta)$;} \mbox{see Fig.\ref{allexampl})}
 as the shower direction tends to the horizon. This
 singularity is the strict consequence of relation (\ref{32}): if
 the ort projection on the horizontal plane approaches unity, the
 reconstructed ``vertical'' component estimation becomes worse evaluated.

 Furthermore, sometimes the ``vertical'' ort
component, evaluated through the nonlinear relation (\ref{32})
with the ``horizontal'' components containing some errors, becomes
imaginary. Really, the estimation
 of ort projection length onto the horizontal plane can prove to be
 greater then unity due to the errors of those components. It is quite
 unclear how one should interpret such result.

 No such accident takes place using the volumetric
EAS goniometer, as all ort components are computed on the base of
the measured delay periods (all being real numbers) by means of
the real linear transformation (\ref{11}). It cannot have a
complex result. Any error can only affect the unit length
condition of the spatial ort \textbf{n}. The linear distortion of
this type never changes its geometrical sense of a direction
vector, and both angles can be calculated quite intelligently,
though possibly with big errors.

\section{Examples}
\vspace{-1.0cm}
\begin{figure}[h]
   \includegraphics[
    natwidth=16cm
   ,natheight=6.15cm
   ,width=16cm
   ,height=6.15cm
   ]
   {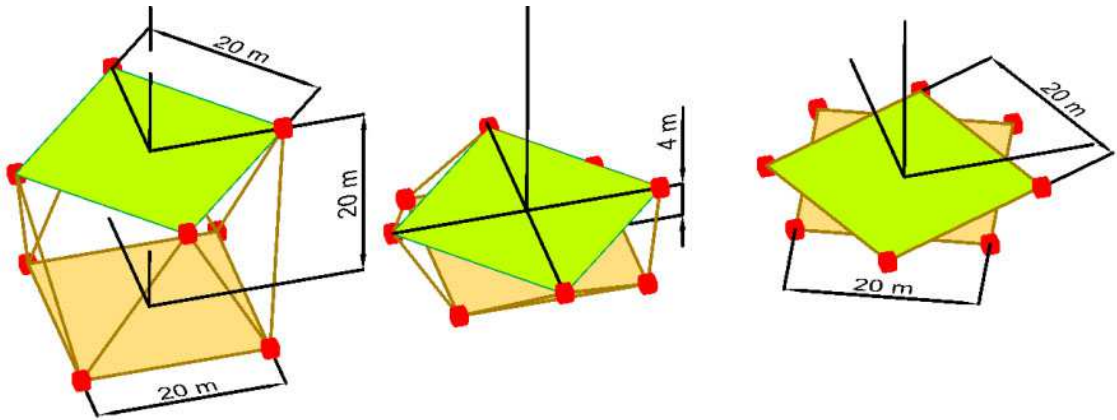}
  \caption
            {\mbox{\quad Three} examples of goniometer arranging:\newline
   \mbox{\qquad 1) \emph{high goniometer} -- left;}
   \quad\mbox{2) \emph{low goniometer} -- center;}
   \quad\mbox{3) \emph{flat goniometer} -- right.}
            }
   \label{exam_3}
   \vspace{0.5cm}
 \end{figure}

 Let us consider three simple examples of EAS goniometers for the
purpose of illustration. (see Fig.\ref{exam_3}) The detector
number in every example installation is \mbox{$N_0=8$} with the
distances between them approximately about \mbox{$20\:{m}$}, as in
example in reference \cite{Ochi}. If we locate them in the apexes
of squares with edges of \mbox{$20\:{m}$} length (the radii of
both tiers are of \mbox{$14.1\:{m}$} in size) and estimate the
delay periods' standard error by a rough value of
 \mbox{$ \sigma=0.5\:m \sim 1.7\:ns$,}
 the aforementioned \mbox{expression (\ref{25})} gives us
 \mbox{$D_{h}=0.0003125$}.

   However, let us locate the tiers:
\begin{enumerate}
    \item \mbox{with the distance} between them  equal
\mbox{to $20\:{m}$} (the so-called \emph{high goniometer}). The
dispersion of the ``vertical'' ort component
\mbox{$D_{z}=D_h=0.0003125$;}
    \item \mbox{with the
distance} between the tiers equal \mbox{to $4\:{m}$} (\emph{low
goniometer} ) and the same configuration of tiers as before. The
dispersion of the ``vertical'' ort component increases:
\mbox{$D_{z}=0.0078125$;}
    \item \mbox{with all
detectors} placed in common horizontal plane (\emph{flat
goniometer} ) at the apexes of a regular octagon with the same
circumcircle radius. As before we obtain \mbox{$D_{h}=0.0003125$.}
(The ``vertical'' component is indefinite.)
\end{enumerate}

\begin{figure}[!ht]
\begin{minipage}[c]{.4\linewidth}
\centering \caption{The errors dependencies of spherical angles
estimations
  on the true zenith angle value for all
 considered cases of the EAS goniometer
 examples.} \label{allexampl}
\end{minipage}%
\hspace{.05\linewidth}
\begin{minipage}[c]{.45\linewidth}
\centering
\includegraphics[natwidth=84mm
   ,natheight=102mm
   ,width=84mm
   ,height=102mm]{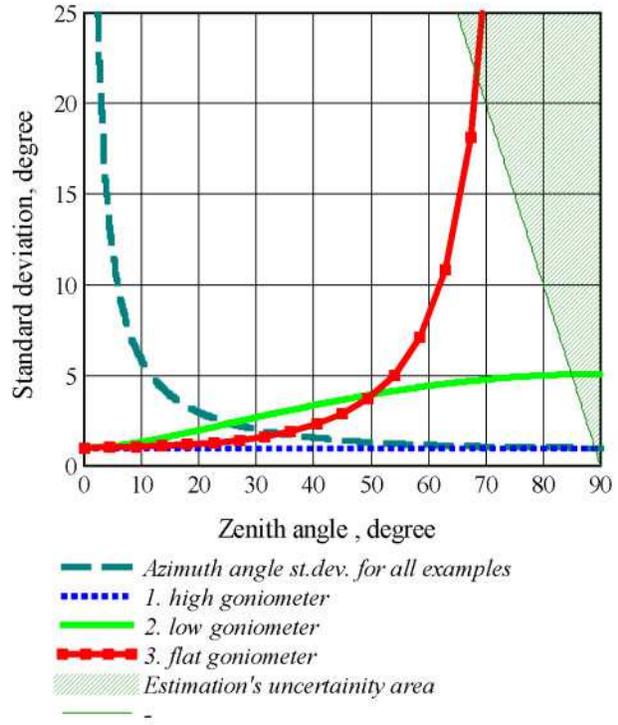}
\end{minipage}
\end{figure}

  The dependencies of angles error estimations on zenith angle value are shown on Fig.\ref{allexampl}
   for all three cases.
 As one can see, the standard deviation of zenith angle estimation
 never exceeds the value of $1^\circ$ for high goniometer; while
 for low goniometer it somewhat exceeds $10^\circ$ limit only for
 almost horizontal showers.
 \begin{figure}[hb]
\begin{minipage}
[c] {.3\linewidth} \centering \caption{The probability of complex
zenith angle estimation.} \label{complexprob}
\end{minipage}%
\hspace{.15\linewidth}
\begin{minipage}[c]{.45\linewidth}
\centering
\includegraphics[width=85mm, height=75mm, natwidth=85mm, natheight=85mm]{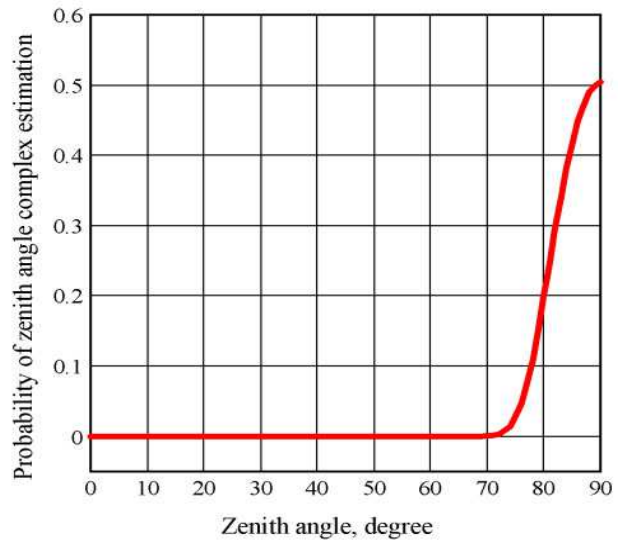}
\end{minipage}
\end{figure}

  However, the LAAS group \cite{Ochi} (together with the most part
 of investigators all over the world) uses the flat goniometer
 installation. All detectors are placed on the same horizontal
 plane preventing the possibility of linear estimation of vertical
 component of EAS direction ort.

      On Fig.\ref{allexampl}  the angle dependencies of error estimations
      for last flat goniometer example are shown, too. The
 azimuth angle is measured with the previous accuracy as the
 detectors' number and their positions radius have not changed.

  The zenith angle estimation error has
 grown badly. Practically every close-to-horizon angles cannot be
 measured as there estimations coincide with the horizon within the
 value of standard deviation. The angles in the shaded area on
 Fig.\ref{allexampl} correspond to this condition. The flat EAS
 goniometer of the last example is not sensitive for zenith angles
 larger then $\sim68^\circ$.

 \begin{figure}[ht]
\begin{minipage}
[c] {.3\linewidth} \centering \caption{Zenith angle estimation by
the flat goniometer:
   \newline \mbox{1) expectation} of the flat goniometer estimation through the real part...
   \newline \mbox{2) expectation} of the desired consistent and unbiased estimation.
   }
\label{expectation}
\end{minipage}%
\hspace{.15\linewidth}
\begin{minipage}[c]{.45\linewidth}
\centering
\includegraphics[width=85mm, height=110mm, natwidth=85mm, natheight=110mm]{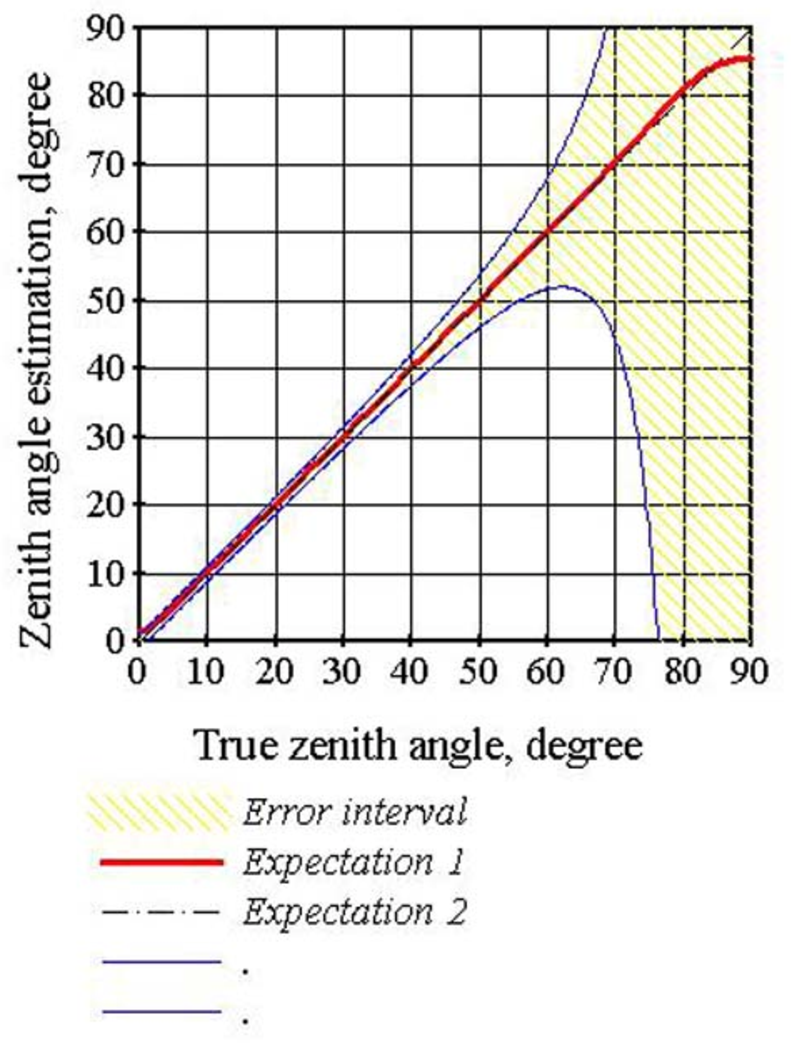}
\end{minipage}
\end{figure}

 The anticipated probability of dummy
 complex estimation of zenith angle is shown on
 Fig.\ref{complexprob} for the flat EAS goniometer considered here,
 as a function of the true zenith angle value. Actually, the
 estimation can become complex in (nearly) the same shaded area on
 Fig.\ref{allexampl}.

 The complexity of the zenith angle estimation indicated,
 caused by the fluctuations of the horizontal components' estimators,
 results in the complexity of the expectation value of the vertical component
 estimator for any value of true zenith angle. The events resulting in complex
 estimation of the vertical component of the EAS directional ort are plainly rejected in practice.
 This means the use of the real part of complete estimator (\ref{32}) for the vertical component estimation.
 This (real) estimator approaches stochastically the real part of the expectation
 value of complete estimator $\Re e \textbf{M}(n_z)$,
 which does not coincide with the true value of the vertical component.
 Hence, this estimator on the trimmed sample is biased and inconsistent one.
 Nevertheless, the use of this value within the interval of angles with negligible
 probability of complex estimations (Fig.\ref{complexprob}) is defensible as the angle's bias value
 does not exceed the bounds of one standard deviation of the resulting zenith angle (Fig.\ref{expectation}).

\section{ Conclusion.}

 \mbox{\qquad The} flat kind of goniometer installation has a
number of unpleasant properties. When the EAS arrival direction
lies far from the zenith, the possibility arises of a complex
estimation of the zenith angle with no clear interpretation. The
standard error of the last angle grows rapidly to infinity for EAS
arrival directions near the horizon. This behavior results in the
assertion of an insistent desirability to only use volumetric EAS
goniometers, especially for EAS network stations with big angular
distances between them. Even a small vertical displacement of part
of the detectors in a flat goniometer array (i.e. conversion to
low EAS goniometer) fundamentally changes the angles computation
conditions: in no case does any complex result arise and the error
in zenith angle proves to have a superior limit even for
horizontal EAS directions. Finally, the proper detectors
arrangement provides the isotropy of errors.

\section{Acknowledgments}

\mbox{\qquad The} authors  are grateful to other current and
former members of our group for their technical support.
     Part of this work was supported by the Georgian National Science Foundation subsidy
     for a grant of scientific researches \#GNSF/ST06/4-075(No~356/07).

 
 %
 \end{document}